\documentclass[letterpaper,10pt,conference]{ieeeconf}

\IEEEoverridecommandlockouts
\usepackage{amsmath}
\usepackage{amssymb}
\usepackage{amsfonts}
\usepackage{amstext}
\usepackage{mathtools}

\usepackage{newtxtext}
\usepackage{newtxmath}
\usepackage[cal=cm]{mathalfa}

\usepackage{graphicx}
\usepackage{epsfig}
\usepackage[table]{xcolor}
\usepackage{booktabs}
\usepackage{float}
\usepackage{caption}
\usepackage{subcaption}
\usepackage{dblfloatfix}

\usepackage{algorithm}
\usepackage{algorithmicx}
\usepackage{algpseudocode}

\usepackage{tikz}
\usetikzlibrary{
    shapes,
    arrows.meta,
    positioning
}

\usepackage[english]{babel}
\usepackage{textcomp}
\usepackage{balance}
\usepackage{comment}
\usepackage{soul}
\usepackage{color}
\usepackage{todonotes}
\usepackage{cite}

\makeatletter
\let\NAT@parse\undefined
\makeatother

\title{
\LARGE \bf
Predictive Rerouting of Connected and Automated Vehicles\\
Using Traffic and Charging Demand Forecasts
}

\author{Tony Kinchen, Ting Bai, and Andreas A. Malikopoulos,~\IEEEmembership{Senior Member, IEEE}
\thanks{This research was supported in part by NSF under Grants CNS-2401007, CMMI-2348381, IIS-2415478, and in part by MathWorks.}\vspace{1.5pt}% <-this % stops a space
\thanks{
Tony Kinchen is with the Systems Engineering Program,
Cornell University, Ithaca, NY 14853, USA. Email:
{\tt\small tjk238@cornell.edu}}%
\thanks{
Ting Bai is with the School of Automation and Intelligent Sensing, Shanghai Jiao Tong University, Shanghai, 200240, China. Email: 
{\tt\small tingbai@sjtu.edu.cn}
}%
\thanks{Andreas A. Malikopoulos is with the Applied Mathematics, Systems Engineering, Robotics, Mechanical Engineering, Electrical and Computer Engineering, and Civil \& Environmental Engineering, Cornell University, Ithaca, NY, USA. (email: \texttt{amaliko@cornell.edu}}%
}

\begin{document}

\maketitle
\thispagestyle{empty}
\pagestyle{empty}

\begin{abstract}
In this paper, we consider the problem of predictive rerouting of
connected and automated vehicles (CAVs) in mixed traffic with electric
vehicle charging demand. We provide a framework that combines a
diffusion convolutional recurrent neural network (DCRNN) with a
routing policy that accounts for congestion, CO$_2$ emissions, route
length, and charging demand. The DCRNN uses historical network
observations to forecast traffic conditions and charging demand.
These forecasts are then used to evaluate feasible alternative
routes for eligible CAVs. A route change is accepted when the
alternative preserves connectivity to the original destination and
improves the prescribed route cost. We evaluate the proposed
framework in SUMO under controlled traffic disruptions at five CAV
penetration levels, ranging from 5\% to 45\%. We compare its
performance with K-shortest-path routing, predictive-density routing,
V2X proactive routing, and a reference scenario without rerouting.
In the considered scenarios, the proposed framework reduces the
average travel-time index by approximately 1.8\% relative to the
reference scenario and achieves the lowest average travel-time
index, highest average speed, and lowest aggregate CO$_2$ emissions
among the active routing methods. Across all five penetration
levels, it accepts 41 route changes, compared with 146 for
K-shortest-path routing and 153 for V2X proactive routing. The
reference scenario retains lower aggregate emissions and distance
traveled, illustrating the tradeoff between congestion reduction
and the additional travel associated with rerouting.
\end{abstract}

%%%%%%%%%%%%%%%%%%%%%%%%%%%%%%%%%%%%%%%%%%%%%%%%%%%%%%%%%%%%%%%%%%%%%%%%%%%
%%%%%%%%%%%%%%%%%%%%%%%%%%%%%%%%%%%%%%%%%%%%%%%%%%%%%%%%%%%%%%%%%%%%%%%%%%%

\section{Introduction}
\label{sec:introduction}

\subsection{Motivation}

Connected and automated vehicles (CAVs) provide an opportunity to
improve transportation network performance by incorporating shared
information into routing decisions. In mixed traffic, however, only
a subset of vehicles can respond to a prescribed routing policy.
The effectiveness of such a policy depends on the information
available when a decision is made, the availability of alternative
routes, and the interactions between rerouted vehicles and the
surrounding traffic. These interactions impose significant
challenges since a route change that benefits an individual vehicle
may increase congestion on other roadway segments.

Transportation electrification introduces additional interactions
between routing decisions and electric vehicle (EV) charging demand.
Vehicles traveling to charging stations contribute to traffic on
the surrounding roadway segments, while congestion affects travel
time, energy consumption, and access to charging infrastructure.
Thus, predicted traffic conditions and charging demand can provide
useful information for selecting alternative routes before a
vehicle reaches an affected area. Incorporating this information
into a routing policy requires evaluating the anticipated benefit
of a route change together with its implications for emissions,
distance traveled, and subsequent traffic conditions.

\subsection{Related Work}

Several research efforts have been reported in the literature
toward developing routing approaches for connected vehicles.
Distributed routing through intelligent intersections has been
investigated as a means of improving throughput and reducing
travel time~\cite{alfaseeh2018distributed}. System-centric routing
has also been considered in mixed traffic, where CAV routing
decisions account for network travel and energy costs while
human-driven vehicles retain individual routing
objectives~\cite{houshmand2019penetration,bang2025integratedrouting}. Other efforts have
developed multi-objective routing approaches that account for
travel time, emissions, fuel consumption, and energy
use~\cite{djavadian2020multiobjective,alfaseeh2020proactive,ahnecorouting,le2024distributed,le2025stochastic}.
These approaches establish the importance of considering the
effects of routing decisions beyond the travel time of an
individual vehicle.

Traffic forecasting provides a means of incorporating anticipated
network conditions into routing decisions. Graph-based models
represent the spatial relationships between roadway segments
together with the temporal evolution of traffic observations.
Spatio-temporal graph convolutional networks, Graph WaveNet,
and adaptive graph convolutional recurrent networks have been
developed for this purpose
\cite{yu2018stgcn,wu2019graphwavenet,bai2020agcrn}.
The diffusion convolutional recurrent neural network (DCRNN)
combines diffusion convolution over a directed graph with
recurrent temporal modeling~\cite{li2018dcrnn}. Extensions using
graph partitioning have also been investigated for traffic
forecasting~\cite{mallick2020dcrnn}. These models provide a basis
for estimating conditions on downstream roadway segments before
a vehicle reaches them. However, forecasting accuracy alone does
not establish the effectiveness of the resulting routing
decisions, since rerouting changes the traffic conditions that
the forecasting model seeks to predict.

EV charging introduces another consideration in the routing
problem. Charging decisions depend on battery state of charge,
travel conditions, station availability, and the time required
to reach and use a charging station. Research efforts on
charging assignment, reinforcement learning, coordinated
charging stops, and integrated transportation and power
networks have examined these interactions
\cite{farhani2023assignment,song2022sumo,schoenberg2021coordinated,liu2022collaborative,grigorev2021electric,bai2025distributedcharging}.
This body of work motivates accounting for charging demand
when evaluating routes through roadway segments that contain
or provide access to charging stations.

Microscopic traffic simulation provides an environment for
evaluating the effects of these decisions on individual vehicles
and the surrounding network. The Simulation of Urban MObility
(SUMO) platform supports the simulation of traffic operations,
vehicle routing, emissions, and EV energy
use~\cite{alvarezlopez2018sumo}. Through its Traffic Control
Interface (TraCI), network observations can be supplied to
forecasting algorithms and routing decisions can be applied
during a simulation. Prior studies have used SUMO to examine
the effects of CAV penetration on traffic performance, fuel
consumption, and emissions
\cite{zhao2022fuel,li2023evaluation}, and to evaluate rerouting
based on connected-vehicle information
\cite{cardenasbenitez2016traffic}.

The above research efforts motivate the integration of traffic
forecasting, charging-demand prediction, and CAV rerouting.
A central question is how predicted conditions can be used to
select route changes that improve traffic performance when
only a subset of vehicles is controllable. Addressing this
question requires evaluating both the information used to
select a route and the realized network performance after
the routing decisions are applied.

The present study differs from purely macroscopic assignment methods because the
route decision is applied to individual vehicles during a microscopic simulation.
This distinction is important for two reasons. First, the feasibility of an
alternative path is determined by the current SUMO edge, the original destination,
and the set of permitted downstream connections, rather than by an aggregate
origin--destination flow alone. Second, the effect of a rerouting action is
realized only after the vehicle interacts with the surrounding mixed traffic.
Consequently, the quality of a candidate route cannot be judged solely by its
standalone length or free-flow travel time. A route that appears attractive for a
single vehicle may place additional demand on another corridor, alter queue
formation near a disruption, or expose the vehicle to edges with elevated
charging-related demand. These considerations motivate a routing policy that uses
predicted local conditions while still evaluating the final outcome through the
microscopic simulation outputs.

The no-reroute reference is also essential for interpreting the results. Active
rerouting can reduce congestion severity by moving controlled vehicles away from
blocked or congested roadway segments, but it may also increase route length and
therefore total distance and emissions. For this reason, we compare the proposed
method both with active routing benchmarks and with the no-reroute reference.
The active-method comparison isolates the relative quality of the routing logic
when interventions are allowed, whereas the no-reroute comparison identifies the
tradeoff introduced by intervention itself. This separation avoids interpreting a
method as universally superior when it improves one metric, such as TTI, while
not improving all aggregate quantities relative to the original routes.

\subsection{Contributions of This Paper}

In this paper, we provide a predictive CAV rerouting framework
that combines DCRNN forecasts with a routing policy implemented
in SUMO. The forecasts are used to evaluate alternative routes
through a cost that accounts for congestion, CO$_2$ emissions,
route length, and exposure to charging demand. For each eligible
CAV, the policy evaluates a finite set of feasible alternatives
to the original destination and applies a route change when
the acceptance conditions are satisfied. Vehicles that are
not eligible for rerouting retain their assigned routes.

The contributions of this paper are:
\begin{enumerate}
    \item An integration of directed graph-based traffic
    forecasting and charging-demand prediction with CAV
    rerouting in a microscopic traffic simulation environment.

    \item A routing procedure that uses predicted conditions
    to rank feasible alternatives and applies connectivity,
    route-cost improvement, and route-change checks when
    accepting a rerouting decision.

    \item A comparative evaluation with K-shortest-path routing,
    predictive-density routing, V2X proactive routing, and a
    reference scenario without rerouting under matched traffic
    demand and disruption schedules at five CAV penetration
    levels.
\end{enumerate}
The evaluation examines the trade-offs among travel-time index,
average speed, emissions, distance traveled, recorded charging
energy, and the number of accepted route changes.

The remainder of the paper is organized as follows.
Section~II presents the simulation setting, forecasting and
routing methods, and performance measures.
Section~\ref{sec:results} presents and discusses the simulation
results. Finally, Section~\ref{sec:conclusion} provides
concluding remarks and directions for future research.

%%%%%%%%%%%%%%%%%%%%%%%%%%%%%%%%%%%%%%%%%%%%%%%%%%%%%%%%%%%%%%%%%%%%%%%%%%%
%%%%%%%%%%%%%%%%%%%%%%%%%%%%%%%%%%%%%%%%%%%%%%%%%%%%%%%%%%%%%%%%%%%%%%%%%%%

\section{Modeling Framework and Predictive Routing}
\label{sec:framework}

This section describes the SUMO network, the controlled disruption
scenarios, the routing policies, and the performance measures used in
this study.

\subsection{Transportation Network and Simulation Setup}

We consider an urban New York City network in SUMO represented by a
directed graph $G=(J,E)$, where $J$ is the set of junctions and $E$ is
the set of directed road edges. A feasible route is an ordered sequence
$R=(e_1,\ldots,e_n)$ whose consecutive edges satisfy SUMO connectivity.
The NYC-6 network contains 638 SUMO edges, including 368 non-internal
road edges, and 254 junctions, including 211 non-internal junctions.
The CAV penetration cases use demand files with 589, 582, 573, 567,
and 558 vehicle or trip entries for the 5\%, 15\%, 25\%, 35\%, and
45\% cases, respectively. The same demand, disruption schedule, and
charging-station configuration are used by all active methods at a
given penetration level. The SUMO configuration files use start time
0 s and step length 1 s; no explicit end time is specified, so runs end
according to trip completion and route-file termination conditions.
Each penetration case uses four charging stations on four station edges.
\vspace*{0.18in}
\begin{table}[H]
\centering
\caption{Simulation setup for the NYC-6 scenarios.}
\label{tab:simulation_setup}
\scriptsize
\setlength{\tabcolsep}{3pt}
\begin{tabular}{@{}p{0.31\columnwidth}p{0.61\columnwidth}@{}}
\toprule
Item & Value \\
\midrule
Network & 638 edges; 368 non-internal road edges \\
Junctions & 254 total; 211 non-internal \\
CAV penetration & 5\%, 15\%, 25\%, 35\%, 45\% \\
Demand entries & 589, 582, 573, 567, 558 \\
SUMO timing & Start 0 s; step length 1 s; no explicit end time \\
Charging stations & 4 stations on 4 station edges \\
Compared policies & No-reroute, K-shortest, predictive density, V2X, DCRNN \\
\bottomrule
\end{tabular}
\end{table}

Let $V$ be the vehicle set and $V_C\subseteq V$ the CAVs eligible for
rerouting. The penetration rate is $p=|V_C|/|V|$. The traffic population
contains passenger vehicles and buses. EV membership $V_E\subseteq V$
is a separate powertrain attribute, so an EV can also be a CAV. CAV
membership controls rerouting eligibility; EV membership controls
battery and charging variables.

\subsection{Disruptions and Reference Conditions}

For each penetration level, scheduled disruption events are indexed by
$q\in\mathcal Q$ and specified by $(E_q,t_q^s,t_q^f,V_q)$, where
$E_q$ is the set of affected edges, $t_q^s$ and $t_q^f$ are the start
and finish times, and $V_q$ contains the vehicles used to create the
disruption. Intersection disruptions are generated by stopping selected
vehicles in junction areas, and rear-end disruptions are generated by
placing a vehicle behind a stopped lead vehicle. The no-reroute
reference uses the same mixed traffic and disruptions but applies no
CAV route intervention.

Travel-time index (TTI) is normalized using an isolated-vehicle
reference in which forced disruption stops are removed and each vehicle
is simulated independently on its assigned route. Let $T^0_{i,e}>0$ be
the isolated reference travel time for vehicle $i$ on edge $e$. In the
extraction code, disrupted-run traversal records are inner-merged with
the isolated reference by vehicle identifier and edge identifier before
TTI is computed. Therefore, an edge introduced by rerouting contributes
to reported TTI only when that same vehicle--edge pair has a positive
isolated-reference traversal time; unmatched records are not assigned
artificial reference times.

\subsection{Common Routing Procedure}

At a decision time $t$, an eligible CAV $i$ has current edge $o_i(t)$,
original destination edge $d_i$, and remaining route $R_i(t)$. For
method $m$, candidate routes are
\begin{equation}
    A_i^m(t)=P_K(o_i(t),d_i;G_m(t))\setminus\{R_i(t)\},
    \label{eq:candidate_set}
\end{equation}
where $P_K$ returns at most $K$ loop-free SUMO-connected routes. Each
candidate $R=(e_1,\ldots,e_n)$ receives cost
\begin{equation}
    J_m(R,t)=\sum_{j=1}^{n}w_m(e_j,t).
    \label{eq:generic_route_cost}
\end{equation}
The selected route minimizes this cost over the feasible set
$F_i^m(t)\subseteq A_i^m(t)$,
\begin{equation}
    R_i^{m,*}(t)\in\arg\min_{R\in F_i^m(t)}J_m(R,t),
    \label{eq:route_selection}
\end{equation}
and is accepted only if $\Delta J_i^m(t)=J_m(R_i(t),t)-J_m(R_i^{m,*}(t),t)>0$
and the cooldown, route-change limit, duplicate-route, and feasibility
checks pass. Ties follow the deterministic order returned by the
candidate-search routine.

All active methods use a one-second decision clock. The K-shortest-path
candidate generator uses $K=3$. The K-shortest-path, predictive-density,
and V2X baselines share TTI trigger 1.10, relative improvement
threshold 0.20, absolute cost-drop threshold 1.50, 60 s cooldown, and
at most six non-forced route changes per CAV. DCRNN uses TTI trigger
1.03, 1 s decision interval, 4 s cooldown, and at most three route
changes per CAV; accepted non-forced DCRNN reroutes must also pass the
DCRNN route-cost, CO$_2$-improvement, and length gates. Generic SUMO
fallback rerouting is disabled for DCRNN.

\subsection{Forecasting and Routing Policies}

\subsubsection{DCRNN-Based Predictive Routing}

The proposed method uses a DCRNN to predict edge-level conditions from
historical observations. Let $X_t\in\mathbb R^{|E|\times n_x}$ be the
edge-feature matrix and $W\in\mathbb R^{|E|\times |E|}$ the directed
adjacency matrix built from permitted downstream SUMO connections after
filtering to the retained road edges. The forecast is
\begin{equation}
    \widehat{\mathcal Y}_t=f_\theta(X_{t-(H-1)\delta_f},\ldots,X_t;W),
    \label{eq:dcrnn_forecast}
\end{equation}
where $H=5$, $P=2$, and $\delta_f=60$ s, yielding a 120 s nominal
forecast horizon. Vertices correspond to the filtered SUMO road edges
and therefore to the rows of $X_t$ and of $W$. The edge features include
TTI, CO$_2$, average speed, collision/spike indicators, EV count,
EV need-charge count and ratio, minimum and mean state of charge, EV
driving energy, nearby EV need-charge count, station indicators, and a
need-charge spike flag. Required targets are TTI and total CO$_2$ mass;
EV need-charge quantities are appended when available. The DCRNN uses
diffusion order 2 and hidden dimension 64 and is trained for up to
450 epochs with Adam, learning rate $7\times10^{-4}$, patience 25, and
an 80/20 chronological train/validation split. The live routing run uses
only feature history available up to the current decision time. The edge graph is
constructed from the SUMO network topology after removing internal movements, so
the graph nodes, feature rows, and routing-cost edge identifiers use the same
filtered edge set. The chronological split and live-history restriction are used
to avoid training or preprocessing choices that would expose the routing policy
to future observations from the evaluation run.

DCRNN assigns edge cost
\begin{equation}
    w_{\mathrm{DCRNN}}(e,t)=aT'_e(t)+bC'_e(t)+cl'_e+dQ_e(t),
    \label{eq:dcrnn_edge_cost}
\end{equation}
with $a=0.95$, $b=3.1$, $c=0.90$, and $d=2.5$. The TTI term is the
DCRNN-predicted TTI at the forecast horizon, normalized by high value
5.0. The CO$_2$ term is the current edge CO$_2$ value, or a length-based
proxy when unavailable, min--max normalized over candidate-edge values.
The length term is normalized edge length. The charging term $Q_e(t)$ is
a dimensionless EV charge-need risk from the nearest predicted time bin
or, when a time-indexed risk is unavailable,
\begin{equation}
    Q_e(t)=\phi_e(t)\max\left\{
    \frac{n_e(t)+\tilde n_e(t)}{\max_{e'}(n_{e'}(t)+\tilde n_{e'}(t))},
    s_e(t)\right\},
\end{equation}
where $n_e$ is predicted EV need-charge count, $\tilde n_e$ is nearby
need-charge count, $s_e$ is the spike flag, and $\phi_e(t)$ downweights
bins far from the estimated arrival time. Thus, the TTI and EV-risk
terms are forecast-based, while CO$_2$ uses current or proxy information.

For accepted DCRNN reroutes, Table~\ref{tab:dcrnn_internal_benefits}
uses raw old-route minus new-route component differences: predicted TTI
component, CO$_2$ route-score proxy, dimensionless EV-risk exposure, and
physical route length in meters. These values explain the route choices
at decision time and are not independent counterfactual measurements of
realized network performance.

The implementation records every accepted route change together with the old and
new candidate routes, the score components used at the decision time, and the
vehicle associated with the decision. These logs provide an audit trail that is
separate from the network-level performance measures. The audit trail is used to
verify that the DCRNN policy is not merely producing more route changes, but is
selecting alternatives that satisfy the route-score improvement and feasibility
requirements. At the same time, the recorded route-score reductions are not used
as substitutes for realized traffic outcomes. Realized outcomes are computed only
from SUMO output files after the simulation run has completed.

The battery-output quantities are treated more cautiously. The recorded EV
battery fields are retained as diagnostic indicators of charging-demand exposure,
but they are not used as the primary evidence for energy-efficiency conclusions.
This distinction is important because some SUMO battery-output fields can behave
as cumulative quantities over time. If such a cumulative field is summed at every
time step, the resulting aggregate can overstate delivered energy. Therefore, EV
battery-output values are reported as recorded diagnostic aggregates, while the
main performance claims are based on TTI, speed, CO$_2$, distance, and rerouting
volume.

\subsubsection{Benchmark Routing Policies}

The K-shortest-path benchmark evaluates up to $K=3$ loop-free candidate
routes using
\begin{equation}
    w_{\mathrm{KSP}}(e,t)=1.2T'_e(t)+1.0C'_e(t)+0.6l'_e,
    \label{eq:ksp_edge_cost}
\end{equation}
where the terms are normalized current congestion, current or proxy
CO$_2$, and edge length. Predictive-density routing uses density
$\rho_e(t)=N_e(t)/l_e$ in vehicles/km and the recent density growth rate
$r_e(t)=(\rho_e(t)-\rho_e(t-\Delta t_\rho))/\Delta t_\rho$, with
$\Delta t_\rho$ normally 1 s. With critical density $\rho_c=80$
vehicles/km, threshold time $T_c=30$ s, penalty gain $\gamma_\rho=0.10$,
and free-flow edge time $w_e^0$, the score is
\begin{equation}
S_e^{\mathrm{PD}}(t)=
\begin{cases}
w_e^0, & \tau_e(t)>T_c,\\
w_e^0+\gamma_\rho(T_c-\max\{0,\tau_e(t)\}), & \tau_e(t)\leq T_c,
\end{cases}
\end{equation}
where $\tau_e(t)$ is the predicted time to critical density. The same
baseline weights 1.2, 1.0, and 0.6 are then applied to congestion,
CO$_2$, and length terms.

The V2X benchmark uses the prescribed disruption schedule with warning
horizon $\Delta t_{\mathrm{V2X}}=180$ s and an eight-edge lookahead. Let
$B(t)=\cup_{q:t_q^s-\Delta t_{\mathrm{V2X}}\leq t\leq t_q^f}E_q$ be the
warning-edge set. A route is considered for proactive rerouting when its
remaining edge set intersects $B(t)$, and admissible bypass candidates
must avoid $B(t)$. V2X uses
$w_{\mathrm{V2X}}(e,t)=1.2T'_e(t)+1.0C'_e(t)+0.6l'_e$. If no
warning-free bypass exists or the best bypass fails the common
acceptance checks, the vehicle keeps its current route, except for the
hard collision-window fallback used to prevent a CAV from entering an
active scripted incident zone.

\subsection{Performance Measures}

Let $\Omega_{m,p}$ be the retained traversal records for method $m$ and
penetration $p$ after removing internal movements, staging-vehicle
records, and unmatched reference records. For record $r$, with duration
$T_r$, distance $d_r$, CO$_2$ mass $c_r$, and reference time
$T^0_{i(r),e(r)}$, the speed and TTI are
\begin{equation}
    v_r=\frac{d_r}{T_r},\qquad
    \mathrm{TTI}_r=\frac{T_r}{T^0_{i(r),e(r)}}.
    \label{eq:tti}
\end{equation}
Average TTI and average speed are arithmetic averages over retained
traversals, while total distance and CO$_2$ are summed over retained
traversals. Speeds are reported in km/h, distances in km, and CO$_2$ in
kg after converting SUMO milligrams by $10^6$.

Accepted route changes are
\begin{equation}
    N_{m,p}^{\mathrm{rr}}=\sum_{i\in V_C}\sum_{t\in\mathcal T_i^{m,p}}
    \mathbf{1}\{R_i(t^+)\neq R_i(t^-)\}.
    \label{eq:reroute_count}
\end{equation}
For the no-reroute reference $m=0$, the percentage TTI reduction is
\begin{equation}
    \eta_{m,p}=100\frac{\overline{\mathrm{TTI}}_{0,p}-
    \overline{\mathrm{TTI}}_{m,p}}{\overline{\mathrm{TTI}}_{0,p}}.
\end{equation}
In Table~\ref{tab:aggregate_method_comparison}, reroutes, CO$_2$,
distance, and diagnostic EV kWh are summed over the five penetration
scenarios, while average reroutes, average TTI, speed, and safety rows
are arithmetic averages over the five scenarios. The reported 1.8\% TTI
improvement is computed from scenario-averaged TTIs. The EV kWh value is
retained only as a diagnostic battery-output aggregate because the audit
indicated that sampled SUMO battery fields may be cumulative or
monotonic; it is not used as an independent energy-efficiency claim.

%%%%%%%%%%%%%%%%%%%%%%%%%%%%%%%%%%%%%%%%%%%%%%%%%%%%%%%%%%%%%%%%%%%%%%%%%%%
%%%%%%%%%%%%%%%%%%%%%%%%%%%%%%%%%%%%%%%%%%%%%%%%%%%%%%%%%%%%%%%%%%%%%%%%%%%

\section{Results}
\label{sec:results}

This section compares K-shortest-path routing, predictive-density
routing, V2X proactive routing, and the proposed DCRNN method at CAV
penetration levels of 5\%, 15\%, 25\%, 35\%, and 45\%. The no-reroute
reference uses the same disrupted traffic conditions without CAV route
intervention. Results are single deterministic runs for each case and
method; no random-seed confidence intervals are reported. Collision
outputs are used only as safety diagnostics.

\subsection{Aggregate Method Comparison}

\begin{table}[H]
\centering
\caption{Aggregate method comparison across all CAV penetration levels.}
\label{tab:aggregate_method_comparison}
\scriptsize
\setlength{\tabcolsep}{1.7pt}
\begin{tabular*}{\columnwidth}{@{\extracolsep{\fill}}lrrrr@{}}
\toprule
Metric & K-Short. & Pred. & V2X & DCRNN \\
\midrule
Reroutes & 146 & 5 & 153 & 41 \\
Avg. reroutes & 29.20 & 1.00 & 30.60 & 8.20 \\
Avg. TTI & 3.570 & 3.591 & 3.582 & 3.520 \\
Speed (km/h) & 27.387 & 27.393 & 27.333 & 27.618 \\
CO$_2$ (kg) & 773.127 & 781.528 & 779.992 & 770.928 \\
Distance (km) & 1566.946 & 1565.129 & 1562.558 & 1563.430 \\
Diag. EV (kWh) & 13470.055 & 13333.564 & 13355.986 & 13367.567 \\
Safety rows & 73.20 & 72.80 & 72.40 & 65.60 \\
\bottomrule
\end{tabular*}
\end{table}

Table~\ref{tab:aggregate_method_comparison} shows that DCRNN achieves
the lowest average TTI, highest average speed, and lowest total CO$_2$
emissions among the four active routing methods. It accepts 41 route
changes across all penetration levels, compared with 146 for
K-shortest-path routing and 153 for V2X. Predictive-density routing is
the most conservative method, with five reroutes and the lowest
diagnostic EV kWh value, but it has the highest average TTI and CO$_2$
among active methods. V2X yields the lowest active-method distance but
requires the largest rerouting volume. These results indicate that DCRNN
improves the congestion--rerouting tradeoff among active methods without
using high-volume route intervention.

\subsection{No-Reroute Reference Comparison}

\begin{figure}[t]
    \centering
    \includegraphics[width=\columnwidth]{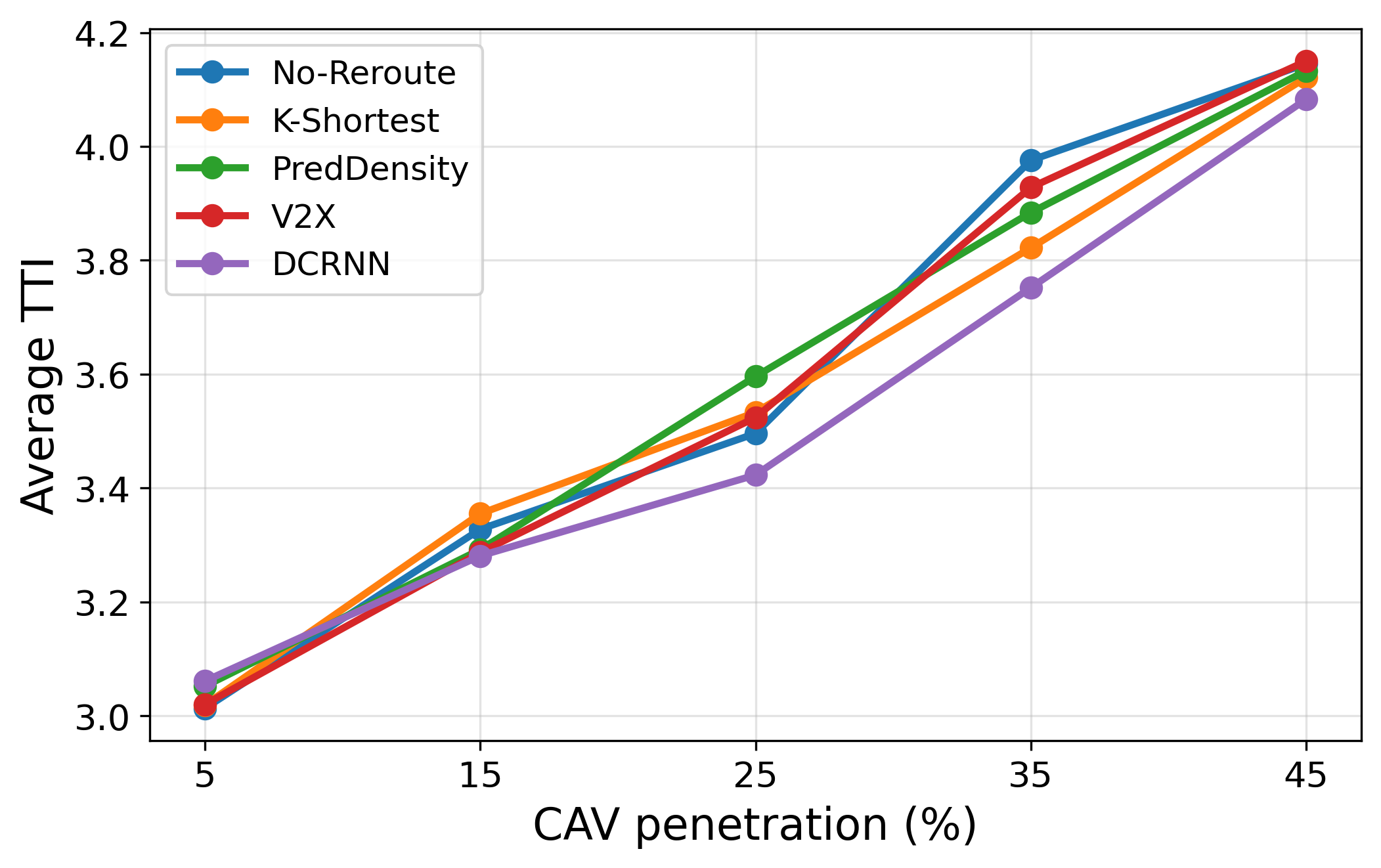}
    \caption{Average TTI across CAV penetration levels with the no-reroute reference.}
    \label{fig:tti_no_reroute}
\end{figure}

Figure~\ref{fig:tti_no_reroute} shows the average TTI trends across
penetration levels. Averaged across the five penetration levels, DCRNN
reduces average TTI by approximately 1.8\% relative to the no-reroute
reference. Thus, the proposed predictive routing method improves the
main congestion-severity metric under the tested disruptions while
remaining selective in the number of accepted route changes.

\begin{figure}[t]
    \centering
    \includegraphics[width=\columnwidth]{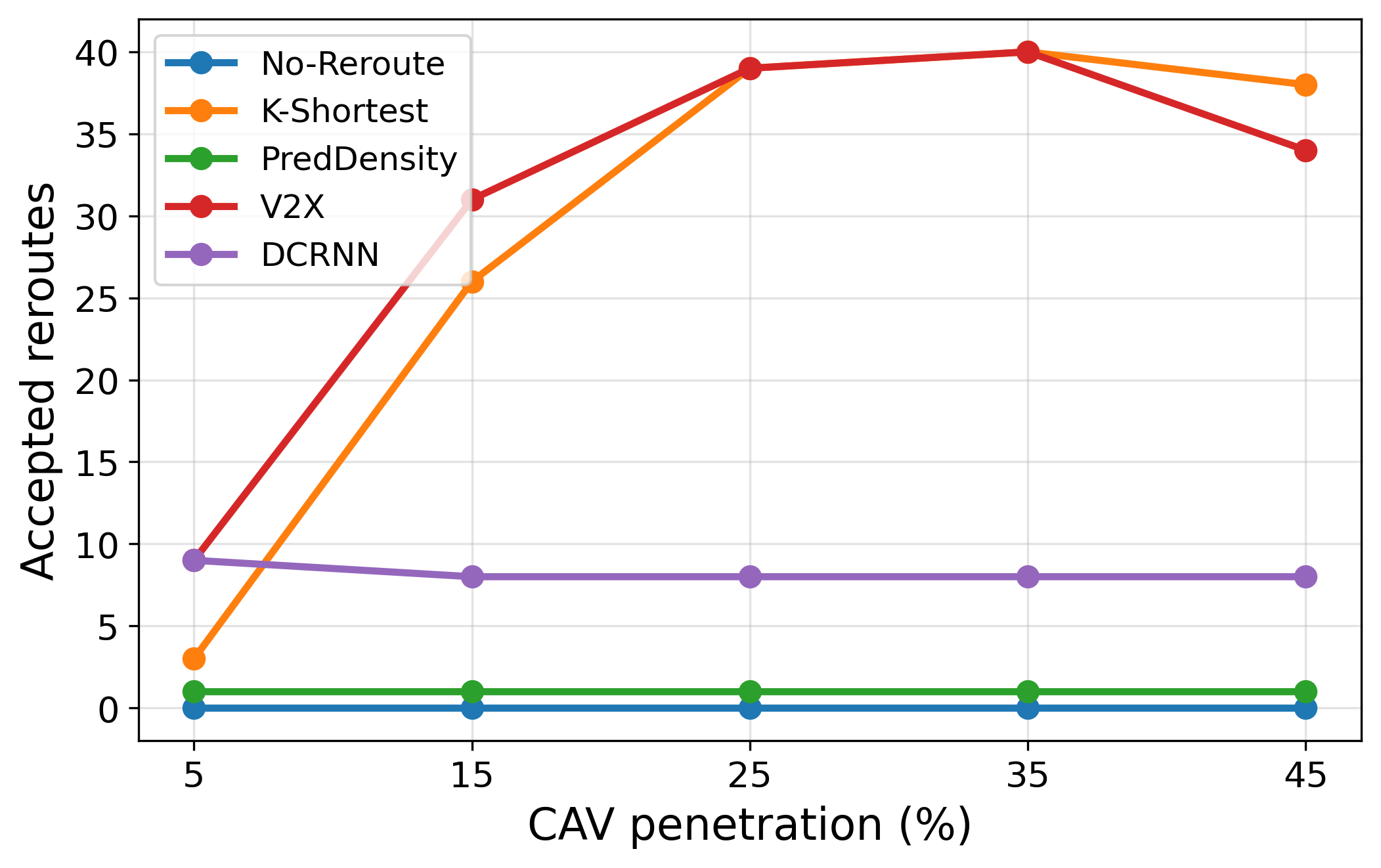}
    \caption{Accepted reroute counts across CAV penetration levels with the no-reroute reference.}
    \label{fig:reroutes_no_reroute}
\end{figure}

Figure~\ref{fig:reroutes_no_reroute} reports the accepted route changes.
The no-reroute reference has zero route changes by definition. K-shortest
and V2X increase their intervention volume as penetration grows, while
DCRNN maintains a small and stable number of accepted reroutes. This
selectivity is important because each unnecessary route change can add
route instability and can move congestion to other parts of the network.
The comparison therefore separates two effects: how often a method
intervenes and whether those interventions improve the realized traffic
state.

\begin{figure}[t]
    \centering
    \includegraphics[width=\columnwidth]{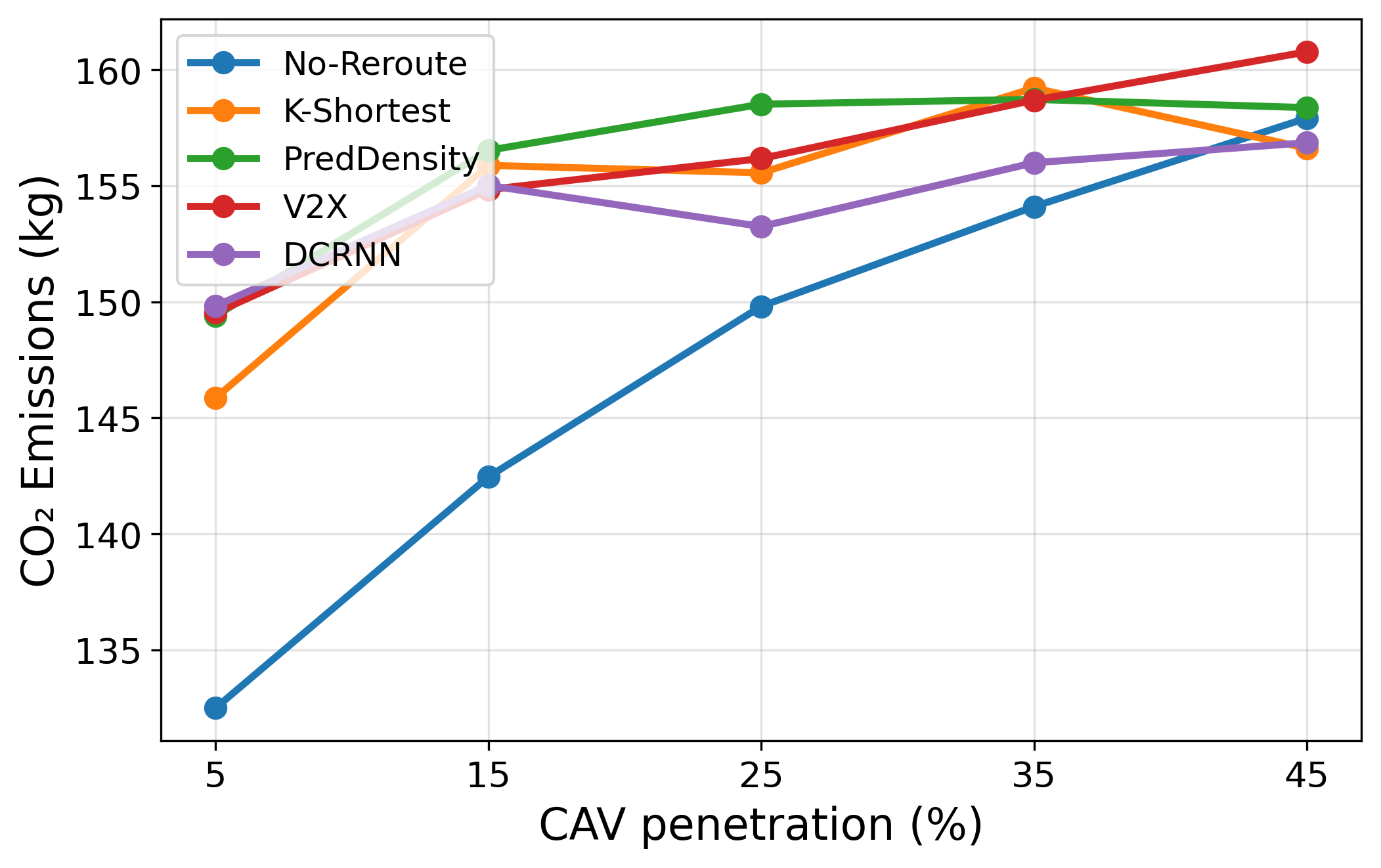}
    \caption{CO$_2$ emissions across CAV penetration levels with the no-reroute reference.}
    \label{fig:co2_no_reroute}
\end{figure}

Figure~\ref{fig:co2_no_reroute} shows that DCRNN has the lowest
aggregate CO$_2$ emissions among active routing methods. However, the
no-reroute reference has lower aggregate CO$_2$ emissions and distance
because vehicles do not deviate from their original routes. Therefore,
the supported CO$_2$ claim is limited to comparisons among active routing
methods, while the strongest no-reroute benefit is the TTI reduction.

\begin{figure}[t]
    \centering
    \includegraphics[width=\columnwidth]{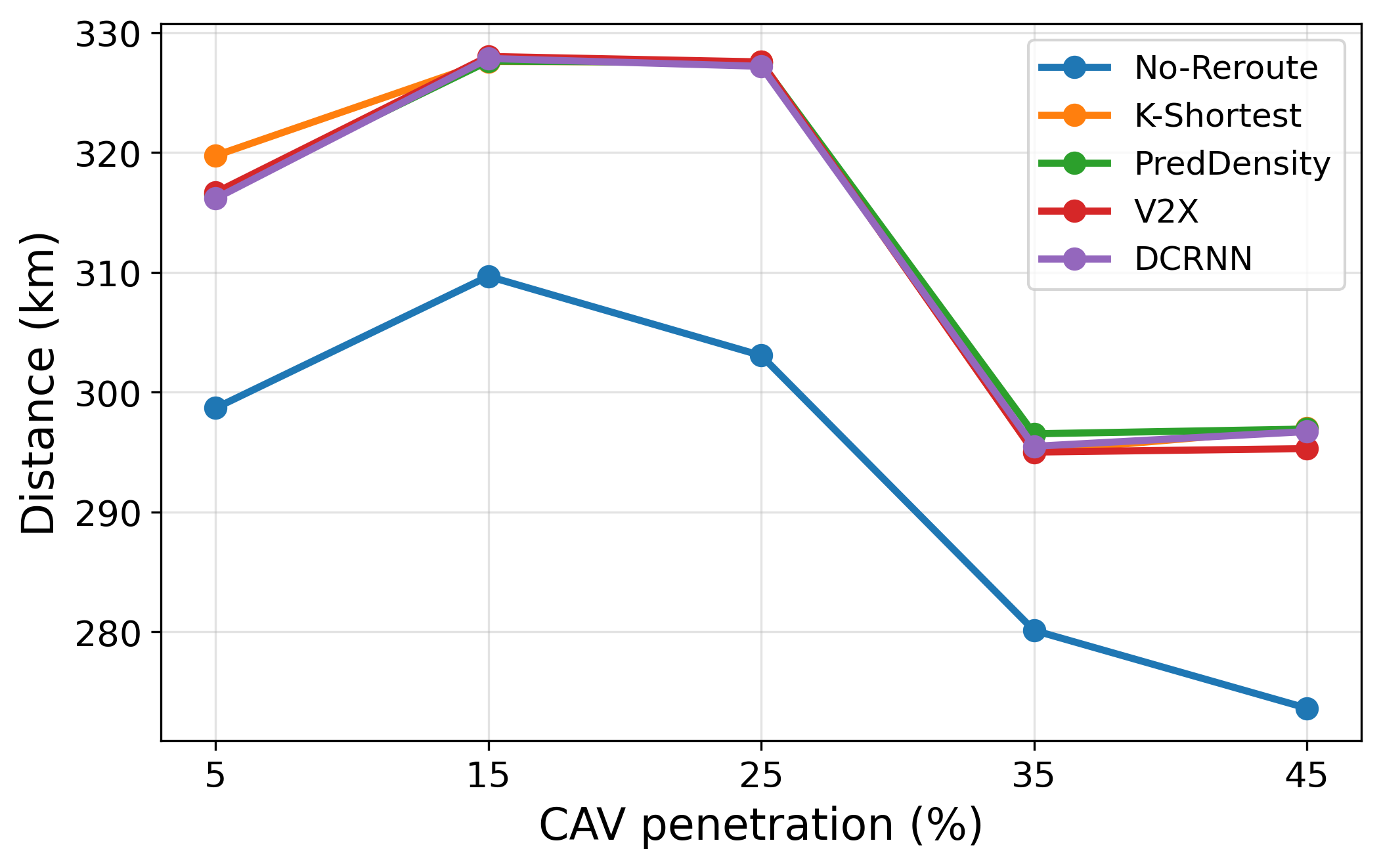}
    \caption{Total distance traveled across CAV penetration levels with the no-reroute reference.}
    \label{fig:distance_no_reroute}
\end{figure}

Figure~\ref{fig:distance_no_reroute} gives the corresponding distance
comparison. The no-reroute reference has the lowest aggregate distance,
which is expected because vehicles remain on their original routes.
Among active methods, V2X has the lowest total distance, while DCRNN is
close to V2X and applies many fewer route changes. These results show
that the proposed method is not a distance-only policy. Instead, DCRNN
balances route length with predicted congestion, CO$_2$, and charging-risk
terms, so the accepted route can be longer when the route score indicates
a congestion benefit.

\begin{figure}[t]
    \centering
    \includegraphics[width=\columnwidth]{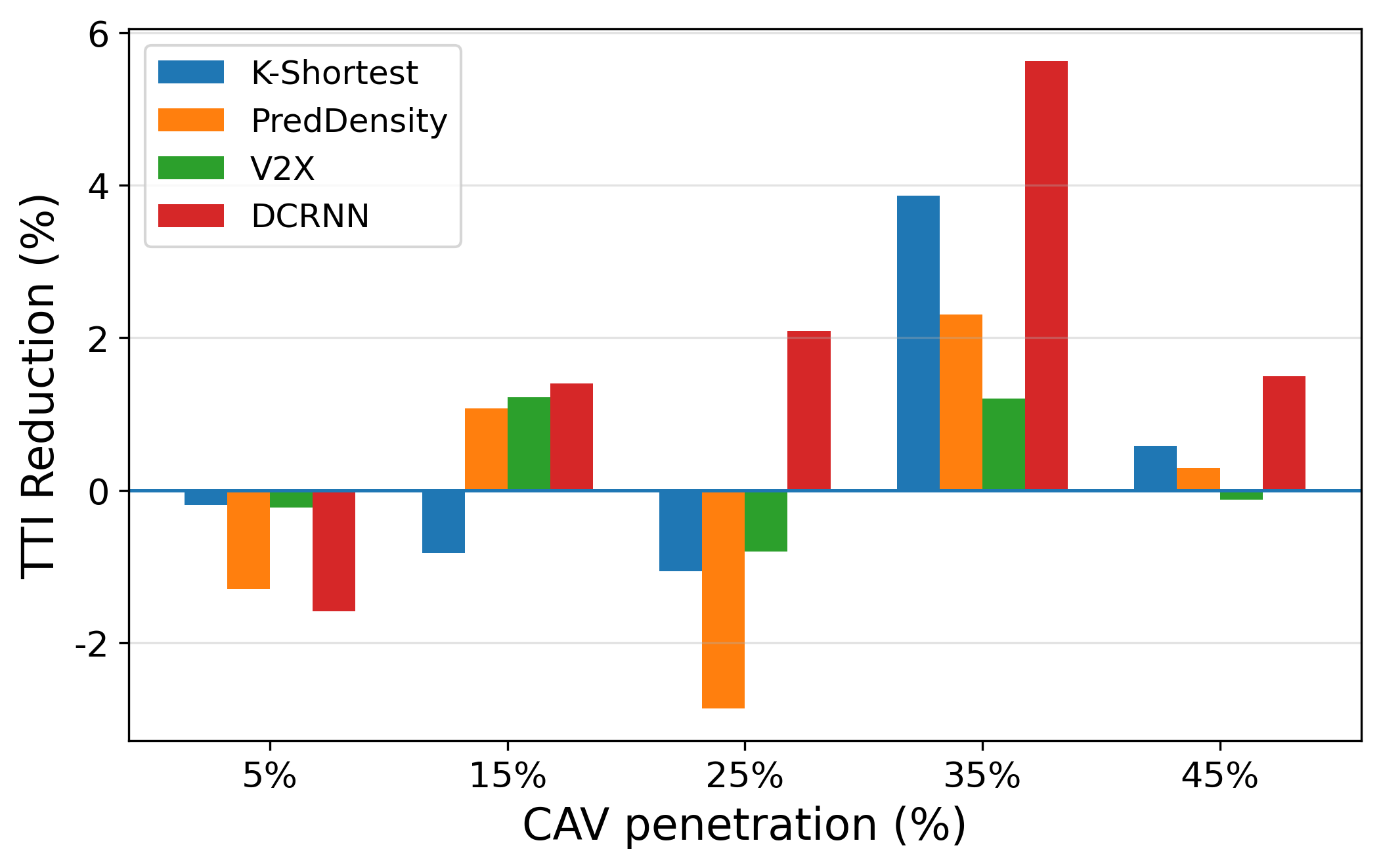}
    \caption{TTI reduction relative to the no-reroute reference. Positive values indicate improvement.}
    \label{fig:tti_reduction_no_reroute}
\end{figure}

Figure~\ref{fig:tti_reduction_no_reroute} reports percentage TTI
reduction relative to no-reroute. The results show clearer positive
DCRNN improvements at higher CAV penetration levels, where more
controllable vehicles are available. Across methods, rerouting volume
alone is not sufficient: predictive-density routing changes very few
routes but does not achieve the best congestion or emissions outcomes,
whereas K-shortest-path and V2X reroute many more vehicles without
matching DCRNN's average TTI or speed. The higher-penetration cases are also
useful for interpreting the role of controllability. As the CAV share increases,
the routing policy has more opportunities to intervene before vehicles reach the
affected areas. However, the results do not show that more control authority
should automatically be converted into more route changes. Instead, the strongest
outcomes occur when the policy uses the larger eligible vehicle set to choose a
small number of higher-value interventions. This observation is consistent with
the accepted-reroute records in Table~\ref{tab:dcrnn_internal_benefits}, where
DCRNN maintains a similar number of accepted reroutes across the 15--45\% cases
while the aggregate route-score reductions increase substantially.

\subsection{Average Multi-Objective Radar Comparison}

The average radar plot in Fig.~\ref{fig:radar_average} provides a compact
summary of active-method tradeoffs. Each spoke is normalized across the
active methods so that 1 denotes the best method on that metric and 0
denotes the weakest method. Lower raw values are preferred for reroutes,
TTI, CO$_2$, distance, and the diagnostic EV battery-output aggregate,
while higher raw speed is preferred. The radar plot is therefore a
relative visualization, not a replacement for the raw values in
Table~\ref{tab:aggregate_method_comparison} and the line plots.

\subsection{Average Multi-Objective Radar Comparison}

\par\medskip
\noindent
\begin{minipage}{\columnwidth}
\centering
\includegraphics[
    width=0.70\columnwidth,
    keepaspectratio
]{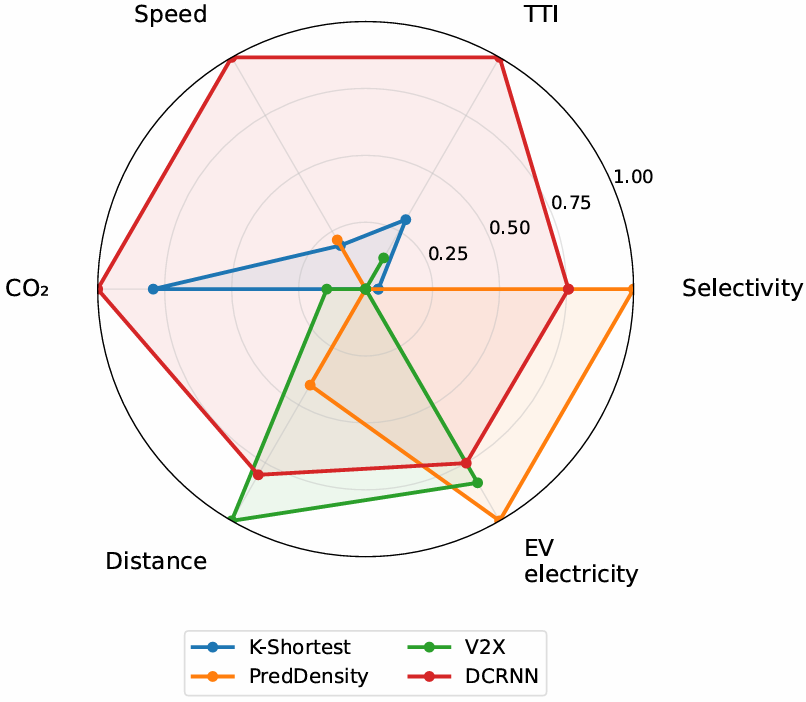}

\captionof{figure}{Average normalized radar comparison.}
\label{fig:radar_average}
\end{minipage}
\par\medskip

DCRNN has the strongest normalized performance for TTI, speed, and
CO$_2$ among the active methods, while retaining better rerouting
selectivity than K-shortest and V2X. Predictive-density dominates the
selectivity and diagnostic EV-kWh spokes because it reroutes very few
vehicles, but its weaker TTI and CO$_2$ outcomes show that low
intervention alone is not sufficient. V2X has the strongest distance
score but requires the highest rerouting volume. This visualization is
consistent with the raw aggregate results: DCRNN provides the most
balanced active-method congestion--rerouting tradeoff, while the
no-reroute reference remains the lower-distance and lower-CO$_2$ case.

\subsection{DCRNN Accepted-Reroute Benefits}

\par\medskip
\noindent
\begin{minipage}{\columnwidth}
\centering
\captionof{table}{DCRNN accepted-reroute objective benefits. Values are computed
from accepted DCRNN reroutes using the old-route and new-route objective fields.}
\label{tab:dcrnn_internal_benefits}
\scriptsize
\setlength{\tabcolsep}{1.6pt}
\resizebox{0.96\columnwidth}{!}{%
\begin{tabular}{@{}lrrrrr@{}}
\toprule
Case &
Reroutes &
TTI red. &
CO$_2$ obj. red. &
EV-risk red. &
Length red. (m) \\
\midrule
CAV 5\%  & 9 &   9.288 &    449.550 & 1.009 & 449.360 \\
CAV 15\% & 8 &  37.037 &  11897.873 & 1.222 & 374.620 \\
CAV 25\% & 8 & 120.841 &  36600.540 & 3.984 & 187.770 \\
CAV 35\% & 8 & 115.650 &  34075.739 & 2.062 & 187.770 \\
CAV 45\% & 8 & 120.859 &  30251.793 & 1.942 & 187.770 \\
\midrule
Total    & 41 & 403.675 & 113275.495 & 10.219 & 1387.290 \\
\bottomrule
\end{tabular}%
}
\end{minipage}
\par\medskip

Table~\ref{tab:dcrnn_internal_benefits} provides an internal analysis
of the accepted DCRNN reroutes. Across all penetration levels, DCRNN
accepted 41 reroutes. Within the DCRNN route-scoring records, these
accepted reroutes were associated with cumulative reductions of
403.675 in the TTI objective, 113275.495 in the CO$_2$ objective,
10.219 in EV charging-risk exposure, and 1387.290 m in route length.
The internal reroute records therefore indicate that the accepted
DCRNN route changes were consistent with the objective components used
by the proposed routing policy. These values should be interpreted as
route-score and candidate-route evidence rather than as independent
counterfactual measurements of the network state. The realized network
comparison is provided by the no-reroute and active-method results above,
whereas Table~\ref{tab:dcrnn_internal_benefits} explains why the DCRNN
policy accepted the specific route changes that it applied. The largest
objective reductions occur in the higher penetration cases, where more eligible
CAVs are available and the same disruption schedule creates more opportunities
for prediction-informed bypasses. The 5\% case has more accepted DCRNN route
changes because two vehicles receive repeated accepted updates, but the total
objective reduction is smaller than in the higher-penetration cases. This
reinforces the interpretation that reroute count alone is not the key measure;
the route-score change associated with each accepted reroute also matters.

The aggregate results should therefore be read as a set of coupled tradeoffs
rather than as a single scalar ranking. Predictive-density routing is useful as a
low-intervention reference: it changes few routes, but its conservative behavior
leaves more vehicles exposed to the disruption-induced congestion. K-shortest-path
routing and V2X proactive routing represent higher-intervention alternatives.
They create many more route changes, but additional route changes do not
necessarily improve network-level outcomes once the new routes interact with the
same mixed-traffic demand. DCRNN occupies an intermediate regime. It performs
more reroutes than predictive-density routing but far fewer than K-shortest-path
and V2X, while achieving the strongest aggregate TTI and speed among the active
methods. This pattern is consistent with the intended role of prediction: it is
most useful when it supports selective acceptance rather than unrestricted
rerouting.

The comparison with no-reroute clarifies the scope of the benefit. Since the
no-reroute reference does not deviate vehicles from their originally assigned
paths, it can retain lower total distance and lower aggregate emissions. This is
not contradictory to the active-method result. The proposed method is intended to
reduce congestion severity under disruptions while limiting unnecessary
intervention. In the reported runs, this is reflected by the 1.8\% average TTI
improvement relative to no-reroute and by the lower TTI, higher speed, and lower
CO$_2$ among active methods. The results do not support a claim that rerouting
strictly dominates the original routes on every aggregate measure.

The safety diagnostic outputs also indicate that CAV rerouting did not
introduce CAV involvement in the scripted disruption incidents. These
diagnostics are useful for verifying that the routing logic avoided
unsafe CAV behavior, but they are not interpreted as the primary
performance objective. The main contribution of the proposed method is
its ability to improve the congestion--rerouting tradeoff through
selective predictive rerouting. The diagnostic nature of these safety rows is
important because the scripted disruption events are imposed to create controlled
congestion and conflict conditions, not to estimate a network-wide crash rate.
Thus, the safety outputs are used to check that the routing logic does not place
CAVs into the constructed incident windows, while the mobility and emissions
metrics are used to evaluate network performance.

Overall, the results support the conclusion that DCRNN provides the
strongest congestion--rerouting tradeoff among the evaluated routing
methods. DCRNN achieves the lowest average TTI, improves average TTI
by approximately 1.8\% relative to no-reroute, and uses substantially
fewer route changes than K-shortest-path and V2X routing. Among the
active routing methods, DCRNN also achieves the lowest aggregate
CO$_2$ emissions and the highest average speed. The no-reroute
reference remains lower in aggregate CO$_2$ and total distance, so the
proposed method is not interpreted as improving every metric relative
to the uncontrolled case.

%%%%%%%%%%%%%%%%%%%%%%%%%%%%%%%%%%%%%%%%%%%%%%%%%%%%%%%%%%%%%%%%%%%%%%%%%%%
%%%%%%%%%%%%%%%%%%%%%%%%%%%%%%%%%%%%%%%%%%%%%%%%%%%%%%%%%%%%%%%%%%%%%%%%%%%

\section{Conclusion}
\label{sec:conclusion}

In this paper, we presented a predictive routing framework
for CAVs in mixed traffic with EV charging demand. The
framework combines DCRNN forecasts with a routing procedure
that evaluates feasible alternatives using congestion,
CO$_2$ emissions, route length, and charging-demand terms.
We evaluated the framework in SUMO under controlled
disruptions at five CAV penetration levels. Across the
considered scenarios, DCRNN yielded the lowest average
TTI, highest average speed, and lowest aggregate CO$_2$
emissions among the active routing methods, while accepting
fewer route changes than K-shortest-path and V2X routing.
The comparison with the no-reroute reference showed an
aggregate TTI improvement, although the benefit varied
across penetration levels. The reference retained lower
aggregate emissions and distance traveled, illustrating
the trade-offs associated with routing intervention.

Future work will extend the evaluation to larger networks
and repeated demand and disruption realizations. We will
also examine the individual contributions of traffic
forecasting and charging-demand information through
controlled comparisons, and incorporate explicit charging
queues and station availability to evaluate the effects
of routing decisions on charging service. Additional work should also separate
prediction quality from routing-policy quality by replaying matched candidate
route sets under alternative forecast sources. Such ablations would clarify when
the observed benefits are caused by improved congestion prediction, by the
multi-objective acceptance rule, or by the interaction between the two. Finally,
future experiments should replace the diagnostic EV battery-output aggregate with
a verified incremental charging-energy measure so that EV energy and charging
service can be evaluated with the same confidence as TTI, speed, CO$_2$, and
distance.

\balance

\bibliographystyle{IEEEtran}
\bibliography{myBib}

@inproceedings{li2018dcrnn,
  title={Diffusion Convolutional Recurrent Neural Network: Data-Driven Traffic Forecasting},
  author={Li, Yaguang and Yu, Rose and Shahabi, Cyrus and Liu, Yan},
  booktitle={International Conference on Learning Representations},
  year={2018}
}

@inproceedings{yu2018stgcn,
title={Spatio-Temporal Graph Convolutional Networks: A Deep Learning Framework for Traffic Forecasting},
author={Yu, Bing and Yin, Haoteng and Zhu, Zhanxing},
booktitle={Proceedings of the Twenty-Seventh International Joint Conference on Artificial Intelligence},
pages={3634--3640},
year={2018},
doi={10.24963/ijcai.2018/505}
}

@inproceedings{alvarezlopez2018sumo,
title={Microscopic Traffic Simulation using SUMO},
author={{'A}lvarez L{'o}pez, Pablo and Behrisch, Michael and Bieker-Walz, Laura and Erdmann, Jakob and Fl{"o}tter{"o}d, Yun-Pang and Hilbrich, Robert and L{"u}cken, Leonhard and Rummel, Johannes and Wagner, Peter and Wie{\ss}ner, Evamarie},
booktitle={2018 21st International Conference on Intelligent Transportation Systems (ITSC)},
pages={2575--2582},
year={2018},
doi={10.1109/ITSC.2018.8569938},
publisher={IEEE}
}

@article{zhao2022fuel,
title={Fuel Consumption and Traffic Emissions Evaluation of Mixed Traffic Flow with Connected Automated Vehicles at Multiple Traffic Scenarios},
author={Zhao, Bin and Lin, Yalan and Hao, Huijun and Yao, Zhihong},
journal={Journal of Advanced Transportation},
volume={2022},
pages={1--14},
year={2022},
doi={10.1155/2022/6345404}
}

@article{li2023evaluation,
title={Evaluation of Fuel Consumption and Emissions Benefits of Connected and Automated Vehicles in Mixed Traffic Flow},
author={Li, Honggang and Li, Hongtao and Hu, Yi and Xia, Tong and Miao, Qiqi and Chu, Jiangwei},
journal={Frontiers in Energy Research},
volume={11},
pages={1207449},
year={2023},
doi={10.3389/fenrg.2023.1207449},
publisher={Frontiers Media SA}
}

@article{cardenasbenitez2016traffic,
title={Traffic Congestion Detection System through Connected Vehicles and Big Data},
author={C{'a}rdenas-Ben{'i}tez, N{'e}stor and Aquino-Santos, Ra{'u}l and Maga{~n}a-Espinoza, Pedro and Aguilar-Velazco, Jos{'e} and Edwards-Block, Arthur and Medina Cass, Aldo},
journal={Sensors},
volume={16},
number={5},
pages={599},
year={2016},
doi={10.3390/s16050599},
publisher={MDPI}
}

@inproceedings{alfaseeh2018distributed,
title={Impact of Distributed Routing of Intelligent Vehicles on Urban Traffic},
author={Alfaseeh, Lama and Djavadian, Shadi and Farooq, Bilal},
booktitle={2018 IEEE International Smart Cities Conference (ISC2)},
pages={1--7},
year={2018},
doi={10.1109/ISC2.2018.8656941},
publisher={IEEE}
}

@inproceedings{houshmand2019penetration,
title={The Penetration Rate Effect of Connected and Automated Vehicles in Mixed Traffic Routing},
author={Houshmand, Arian and Wollenstein-Betech, Salom{'o}n and Cassandras, Christos G.},
booktitle={2019 IEEE Intelligent Transportation Systems Conference (ITSC)},
pages={1755--1760},
year={2019},
doi={10.1109/ITSC.2019.8916938},
publisher={IEEE}
}

@article{djavadian2020multiobjective,
title={Multi-Objective Eco-Routing for Dynamic Control of Connected \& Automated Vehicles},
author={Djavadian, Shadi and Tu, Ran and Farooq, Bilal and Hatzopoulou, Marianne},
journal={Transportation Research Part D: Transport and Environment},
volume={87},
pages={102513},
year={2020},
doi={10.1016/j.trd.2020.102513},
publisher={Elsevier}
}

@article{alfaseeh2020proactive,
title={Deep Learning Based Proactive Multi-Objective Eco-Routing Strategies for Connected and Automated Vehicles},
author={Alfaseeh, Lama and Farooq, Bilal},
journal={Frontiers in Future Transportation},
volume={1},
pages={594608},
year={2020},
doi={10.3389/ffutr.2020.594608},
publisher={Frontiers Media SA}
}

@article{ahnecorouting,
title={Multi-Objective Eco-Routing Model Development and Evaluation for Battery Electric Vehicles},
author={Ahn, Kyoungho and Bichiou, Youssef and Farag, Mohamed and Rakha, Hesham A.},
journal={Transportation Research Record},
volume={2675},
number={12},
pages={867--879},
year={2021},
doi={10.1177/03611981211031529},
publisher={SAGE Publications}
}

@inproceedings{wu2019graphwavenet,
title={Graph WaveNet for Deep Spatial-Temporal Graph Modeling},
author={Wu, Zonghan and Pan, Shirui and Long, Guodong and Jiang, Jing and Zhang, Chengqi},
booktitle={Proceedings of the Twenty-Eighth International Joint Conference on Artificial Intelligence},
pages={1907--1913},
year={2019},
doi={10.24963/ijcai.2019/264},
publisher={International Joint Conferences on Artificial Intelligence Organization}
}

@inproceedings{bai2020agcrn,
title={Adaptive Graph Convolutional Recurrent Network for Traffic Forecasting},
author={Bai, Lei and Yao, Lina and Li, Can and Wang, Xianzhi and Wang, Can},
booktitle={Advances in Neural Information Processing Systems},
volume={33},
pages={17804--17815},
year={2020},
publisher={Curran Associates, Inc.}
}

@article{mallick2020dcrnn,
title={Graph-Partitioning-Based Diffusion Convolutional Recurrent Neural Network for Large-Scale Traffic Forecasting},
author={Mallick, Tanwi and Balaprakash, Prasanna and Rask, Eric and Macfarlane, Jane},
journal={Transportation Research Record},
volume={2674},
number={9},
pages={473--488},
year={2020},
doi={10.1177/0361198120930010},
publisher={SAGE Publications}
}

@article{farhani2023assignment,
title={Assignment Approach for Electric Vehicle Charging Using Traffic Data Collected by SUMO},
author={Farhani, Riham and El Hillali, Yassin and Rivenq, Atika and Boughaleb, Yahia and Hajjaji, Abdelowahed},
journal={World Electric Vehicle Journal},
volume={14},
number={2},
pages={40},
year={2023},
doi={10.3390/wevj14020040},
publisher={MDPI}
}

@article{song2022sumo,
title={A SUMO Framework for Deep Reinforcement Learning Experiments Solving Electric Vehicle Charging Dispatching Problem},
author={Song, Yaofeng and Zhao, Han and Luo, Ruikang and Huang, Liping and Zhang, Yicheng and Su, Rong},
journal={arXiv preprint arXiv:2209.02921},
year={2022},
doi={10.48550/arXiv.2209.02921}
}

@inproceedings{schoenberg2021coordinated,
title={Coordinated Electric Vehicle Re-Charging to Reduce Impact on Daily Driving Schedule},
author={Schoenberg, Sven and Buse, Dominik S. and Dressler, Falko},
booktitle={2021 IEEE Intelligent Vehicles Symposium (IV)},
pages={1180--1187},
year={2021},
doi={10.1109/IV48863.2021.9575517},
publisher={IEEE}
}

@article{liu2022collaborative,
title={Collaborative EV Routing and Charging Scheduling With Power Distribution and Traffic Networks Interaction},
author={Liu, Jiayan and Lin, Gang and Huang, Sunhua and Zhou, Yang and Rehtanz, Christian and Li, Yong},
journal={IEEE Transactions on Power Systems},
volume={37},
number={5},
pages={3923--3936},
year={2022},
doi={10.1109/TPWRS.2022.3142256},
publisher={IEEE}
}

@inproceedings{grigorev2021electric,
title={How Will Electric Vehicles Affect Traffic Congestion and Energy Consumption: An Integrated Modelling Approach},
author={Grigorev, Artur and Mao, Tuo and Berry, Adam and Tan, Joachim and Purushothaman, Loki and Mihaita, Adriana-Simona},
booktitle={2021 IEEE International Intelligent Transportation Systems Conference (ITSC)},
pages={1635--1642},
year={2021},
doi={10.1109/ITSC48978.2021.9564561},
publisher={IEEE}
}

@article{le2024distributed,
  title={Distributed Optimization for Traffic Light Control and Connected Automated Vehicle Coordination in Mixed-Traffic Intersections},
  author={Le, Viet-Anh and Malikopoulos, Andreas A.},
  journal={IEEE Control Systems Letters},
  volume={8},
  pages={2721--2726},
  year={2024},
  doi={10.1109/LCSYS.2024.3512332},
  publisher={IEEE}
}

@article{le2025stochastic,
  title={Stochastic Time-Optimal Trajectory Planning for Connected and Automated Vehicles in Mixed-Traffic Merging Scenarios},
  author={Le, Viet-Anh and Chalaki, Behdad and Tzortzoglou, Filippos N. and Malikopoulos, Andreas A.},
  journal={IEEE Transactions on Control Systems Technology},
  volume={33},
  number={4},
  pages={1403--1417},
  year={2025},
  doi={10.1109/TCST.2024.3433206},
  publisher={IEEE}
}

@article{bai2025distributedcharging,
  title={Distributed Charging Coordination for Electric Trucks Under Limited Facilities and Travel Uncertainties},
  author={Bai, Ting and Li, Yuchao and Malikopoulos, Andreas A. and Johansson, Karl Henrik and M{\aa}rtensson, Jonas},
  journal={IEEE Transactions on Intelligent Transportation Systems},
  volume={26},
  number={7},
  pages={10278--10294},
  year={2025},
  doi={10.1109/TITS.2025.3550035},
  publisher={IEEE}
}

@article{bang2025integratedrouting,
  title={Optimal Trajectory Planning Meets Network-Level Routing: Integrated Control Framework for Emerging Mobility Systems},
  author={Bang, Heeseung and Malikopoulos, Andreas A.},
  journal={Automatica},
  volume={179},
  pages={112389},
  year={2025},
  doi={10.1016/j.automatica.2025.112389},
  publisher={Elsevier}
}

\end{document}